\documentclass[preprint2]{aastex631}
\begin{document}
\title{The Gaia parallax discrepancy for the cluster Pismis 19, and separating $\delta$ Scutis from Cepheids}

\author[0000-0001-8803-3840]{Daniel Majaess}
\affiliation{Mount Saint Vincent University, Halifax, Canada}
\email{Daniel.Majaess@msvu.ca}

\author[0000-0002-4102-1751]{Charles J.~Bonatto}
\affiliation{Departamento de Astronomia, Universidade Federal do Rio Grande do Sul, CP 15051, RS, Porto Alegre 91501-970, Brazil}

\author[0000-0003-1184-1860]{David G.~Turner}
\affiliation{Saint Mary's University, Halifax, Canada}

\author[0000-0001-6878-8648]{Roberto K. Saito}
\affiliation{Departamento de Física, Universidade Federal de Santa Catarina, Trindade 88040-900, Florianópolis, Brazil}

\author[0000-0002-7064-099X]{Dante Minniti}
\affiliation{Instituto de Astrofísica, Dep.~de Física y Astronomía, Facultad de Ciencias Exactas, Universidad Andres Bello, Av.~Fernández Concha 700, Santiago, Chile}
\affiliation{Vatican Observatory, Specola Vaticana, V-00120, Vatican City, Vatican City State}

\author[0009-0004-3783-6378]{Christian Moni Bidin}
\affiliation{Instituto de Astronomía, Universidad Católica del Norte, Av. Angamos 0610, Antofagasta, Chile}

\author[0000-0003-1698-4924]{Danilo González-Díaz}
\affiliation{Instituto de Astronomía, Universidad Católica del Norte, Av. Angamos 0610, Antofagasta, Chile}
\affil{Universidad de Antioquia, Calle 70 52-21, Medellín, Colombia}

\author[0000-0003-3496-3772]{Javier Alonso-Garcia}
\affiliation{Centro de Astronomía, Universidad de Antofagasta, Av.~Angamos 601, Antofagasta, Chile}
\affiliation{Millennium Institute of Astrophysics, Nuncio Monseñor Sotero Sanz 100, Of.~104, Providencia, Santiago, Chile}

\author[0000-0002-4896-8841]{Giuseppe Bono}
\affil{Department of Physics, Università di Roma Tor Vergata, via della Ricerca Scientifica 1, 00133, Rome, Italy}
\affiliation{INAF - Osservatorio Astronomico di Roma, Via di Frascati 33, 00078 Monte Porzio Catone, Italy}

\author[0000-0003-1184-1860]{Vittorio F. Braga}
\affiliation{INAF - Osservatorio Astronomico di Roma, Via di Frascati 33, 00078 Monte Porzio Catone, Italy}

\author[0000-0002-4896-8841]{Maria G. Navarro}
\affiliation{INAF - Osservatorio Astronomico di Roma, Via di Frascati 33, 00078 Monte Porzio Catone, Italy}

\author[0000-0002-0155-9434]{Giovanni Carraro}
\affiliation{Dipartimento di Fisica e Astronomia “Galileo Galilei,” Università degli Studi di Padova, Vicolo Osservatorio 3, I-35122, Padova, Italy}

\author[0000-0002-4430-9427]{Matias Gomez}
\affiliation{Instituto de Astrofísica, Dep.~de Física y Astronomía, Facultad de Ciencias Exactas, Universidad Andres Bello, Av.~Fernández Concha 700, Santiago, Chile}

\begin{abstract}
Pre-Gaia distances for the open cluster Pismis 19 disagree with Gaia parallaxes.  A 2MASS $JK_s$ red clump distance was therefore established for Pismis 19 ($2.90\pm0.15$ kpc), which reaffirms that zero-point corrections for Gaia are required (e.g., Lindegren et al.~2021). OGLE GD-CEP-1864 is confirmed as a member of Pismis 19 on the basis of DR3 proper motions, and its 2MASS+VVV color-magnitude position near the tip of the turnoff.  That $0^{\rm d}.3$ variable star is likely a $\delta$ Scuti rather than a classical Cepheid.  The case revealed a pertinent criterion to segregate those two populations in tandem with the break in the Wesenheit Leavitt Law ($\simeq 0^{\rm d}.5$). Just shortward of that period discontinuity are $\delta$ Scutis, whereas beyond the break lie first overtone classical Cepheids mostly observed beyond the first crossing of the instability strip.
\end{abstract}


\section{Introduction}
\citet[][]{so20} and \citet[][]{ri23} classified OGLE GD-CEP-1864 as a $0^{\rm d}.3$ first overtone classical Cepheid on the basis of OGLE and Gaia observations. Subsequently, \citet{ri23} and \citet{mt24b} suggested OGLE GD-CEP-1864 could be a Cepheid member of Pismis 19. The putative case is enticing given OGLE GD-CEP-1864 may be the shortest-period cluster Cepheid, thereby providing a critical calibrator for Cepheid period-age relations \citep[e.g.,][]{bo05,tu12,and16}. Such Cepheids are likewise key to constraining the minimum mass sampled by the hot extent of the blue loop at a given chemical composition \citep[e.g.,][]{bo00}.  More broadly, cluster Cepheids are important owing to their use in assessing the validity of the Gaia zero-point, and benchmarking the Planck $\Lambda$CDM value of $H_0$ \citep[e.g.,][]{ra23}.  However, \citet[][CCHP]{fr24d} cautioned that Cepheids produce nearer extragalactic distances than TRGB and JAGB stars, whereas \citet[][S$H_0$ES]{ri24} express a separate view.

The cluster parameters and variable star classification must be confirmed to exploit the connection between Pismis 19 and OGLE GD-CEP-1864. Indeed, the distance to Pismis 19 is contested.  \citet{po21} relied on (E)DR3 to establish a distance to Pismis 19 of 3.5 kpc ($\pi=0.282\pm0.061$ mas), while \citet{cg20} utilized DR2 data to determine a parallax and color-magnitude diagram distance of 3.9 kpc ($\pi=0.255\pm0.086$ mas) and 3.5 kpc, accordingly.  Conversely, \citet{maj12} summarized pre-Gaia distances which converged near 2.5 kpc.  Specifically, \citet{pi98}, \citet{ca11}, and \citet{maj12} obtained cluster distances of $2.40 \pm 0.88$ kpc, $2.5 \pm 0.5$ kpc, and $2.40 \pm 0.15$ kpc, respectively.  The results imply that the Gaia distance to Pismis 19 is too remote, and a correction is needed.   

In this study a near-infrared red clump distance is evaluated for Pismis 19 by drawing upon 2MASS $JHK_s$ photometry and Gaia proper motions \citep{cut03,gaia23}.  Moreover, new VVV photometry\footnote{Near $K_s\sim14^{m}.5$ the formal VVV uncertainties are approximately a tenth those associated with 2MASS.} \citep[][Alonso-Garcia et al.~2025, \textit{in prep.}]{sai12} is used in tandem with 2MASS to construct a color-magnitude diagram, and clarify the evolutionary status of OGLE GD-CEP-1864.  Infrared $JHK_s$ data are used to reduce uncertainties stemming from the extinction law. For example, \citet{car13} noted that the optical ratio $A_V/E(B-V)$ is anomalous toward Westerlund 2 \citep[Carina, see also][]{tur12}, whereas extinction law variations are relatively less pronounced in the infrared \citep[e.g.,][]{maj16}.  The emerging results fostered pertinent constraints on Pismis 19 and OGLE GD-CEP-1864.

\begin{deluxetable}{ccc}
\tablecaption{Pismis 19 distance.\label{table:distances}}
\tablehead{\colhead{Reference} & \colhead{Method} & \colhead{d (kpc)}}
\startdata
P98 & $BVI_c$ & $2.40 \pm 0.88$  \\
C11 & $UBVRI_c$ & $2.5 \pm 0.5$ \\
M12 & $JHK_s$ & $2.40 \pm 0.15$ \\
\hline
This work & RC & $2.90\pm0.15$ \\
\hline
C20 & DR2 & $3.9$ \\
P21 & EDR3 & $3.5$ \\
\enddata
\tablenotetext{ }{Notes:~references are \citet[][P98]{pi98}, \citet[][C11]{ca11}, \citet[][M12]{maj12}, \citet[][C20]{cg20}, \citet[][P21]{po21}.}
\end{deluxetable}

\begin{figure}[t]
\begin{center}
 \includegraphics[width=3.5in]{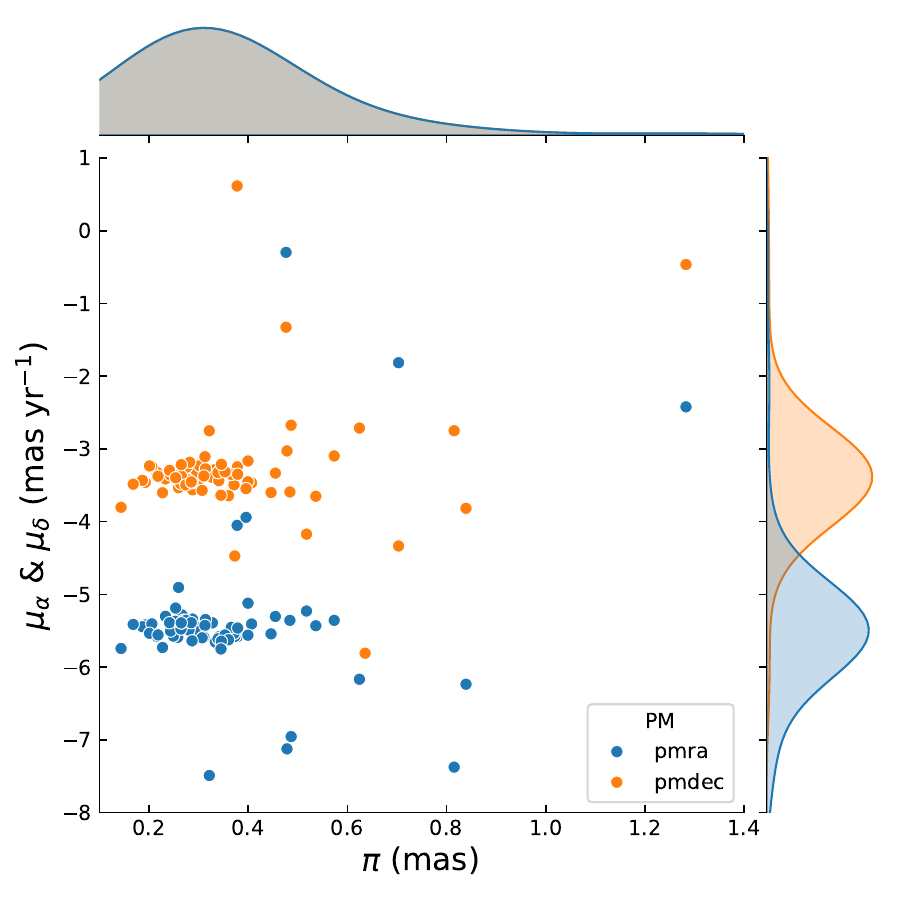} 
  \caption{DR3 astrometry for stars within $\lesssim 0.5 \arcmin$ of the cluster center (Pismis 19).  The overdensities represent probable cluster members ($\mu_{\alpha}$ is blue, while $\mu_{\delta}$ is orange).}
 \label{fig-pm}
\end{center}
\end{figure}

\section{Analysis}
\subsection{Distance to Pismis 19}
An inspection of Gaia DR3 astrometry within $0.5 \arcmin$ of the cluster center reveals clear overdensities (Fig.~\ref{fig-pm}, $n=75$).  Members adhere to the following proper motion:~$\mu_{\alpha}, \mu_{\delta}=-5.48\pm0.11,-3.37\pm0.12$ mas/yr, where the uncertainty is the median absolute deviation (i.e., to reduce the impact of outliers). The parallax is $\pi=0.31\pm0.06$ mas ($d\simeq3.23$ kpc), and places the cluster closer than the \citet{cg20} estimate.  The aforementioned sampling radius was imposed to mitigate field star contamination, and culling criteria were applied (e.g., RUWE$<2$, RPlx$>2$). OGLE GD-CEP-1864 possesses a DR3 parallax of $\pi=0.28\pm0.03$ mas, and the proper motion data are $\mu_{\alpha},\mu_{\delta}=-5.59\pm0.03,-3.30\pm0.04$ mas/yr.  Consequently, OGLE GD-CEP-1864 is a member of Pismis 19. 

The Gaia radial velocity (RV) for OGLE GD-CEP-1864 is $-34.2\pm10.2$~km~s$^{-1}$, where the uncertainty is ostensibly large because of the star’s variable nature. Expanding the search to $r=2\arcmin$ reveals 34 stars with Gaia velocities. Half that sample remains after restricting the analysis to stars with astrometry compatible with cluster membership ($\pi$, $\mu_{\alpha}$, $\mu_{\delta}$), and applying 2$\sigma$-clipping. That subsample of cluster members has a mean $RV=-30.5\pm0.9$~km~s$^{-1}$, which bolsters the association between OGLE GD-CEP-1864 and Pismis 19. 

\begin{figure}[t]
\begin{center}
 \includegraphics[width=2.6in]{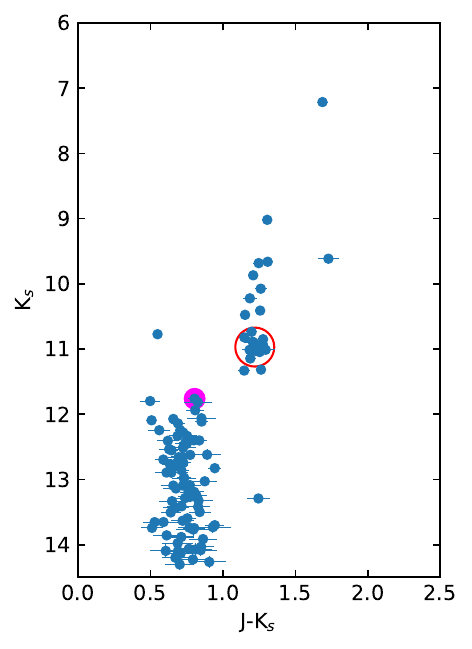} 
  \caption{Probable cluster members (Gaia-cleaned) possessing AAA 2MASS photometry and uncertainties.  The red clump stars are encompassed by a red circle.  The magenta datum represents OGLE GD-CEP-1864.}
 \label{fig-rc}
\end{center}
\end{figure}

A Gaia-cleaned 2MASS color-magnitude diagram is shown as Fig.~\ref{fig-rc} (n=106).  Cluster members were identified by their astrometry within $2\sigma$ of the results derived above, and were restricted to those stars which, for example, possess a 2MASS AAA quality flag.  The sampling radius was expanded to $\simeq 2 \arcmin$ to ensure adequate red clump statistics.  The encircled red clump is readily discernible in the color-magnitude diagram (Fig.~\ref{fig-rc}), and is characterized by $\overline{K_s}=10.97\pm0.11$ and $\overline{J-K_s}=1.22\pm0.04$.  Such findings, in concert with red clump absolute magnitudes and an extinction law \citep[][]{maj11,maj16}, yield $d=2.80\pm0.16$ kpc.  That determination stemmed from a Monte Carlo simulation that included uncertainties associated with each term.  Employing the \citet{lan12} absolute magnitudes ($M_J=-0.984\pm0.014$, $M_{K_s}=-1.613\pm0.015$) results in $d=2.90\pm 0.15$ kpc. The latter estimate is favored since the absolute magnitudes were determined by an independent research team \citep{lan12}.

An attempt was made to mitigate human bias, since a subset of the authors published distances for Pismis 19.  The author R.K.S. pursued a separate derivation of the red clump distance, and utilized the \citet{rd18} absolute magnitudes for red clump giants (i.e., $M_{K_s}=-1.605$, $(J-K_s)_o=0.66$), and extinction laws from \citet{sa22} and \citet{mi18} ($A_{K_s}/E(J-K_s)=0.448,0.484$). Their determined distance supports the $2.9$ kpc finding ($\mu_0=12.331\pm0.204$).  Ideally, a preferable approach to reduce bias is a blind process \citep[e.g.,][CCHP and DESI]{fr24d,ch24}.

The red clump distance, together with the pre-Gaia results, indicate that the DR3 parallaxes for Pismis 19 members are too small, and a zero-point problem exists \citep[e.g.,][]{lin21,ow22}.  Applying the \citet{lin21} correction to stars highlighted in Fig.~\ref{fig-rc} revises the DR3 distance to $\simeq 3.1$ kpc.   Alternatively, a weighted mean was derived employing the \citet{BailerJones21} approach (3.16$\pm$0.10~kpc), which likewise relies on the \citet{lin21} correction. 

\begin{figure}[t]
\begin{center}
 \includegraphics[width=2.6in]{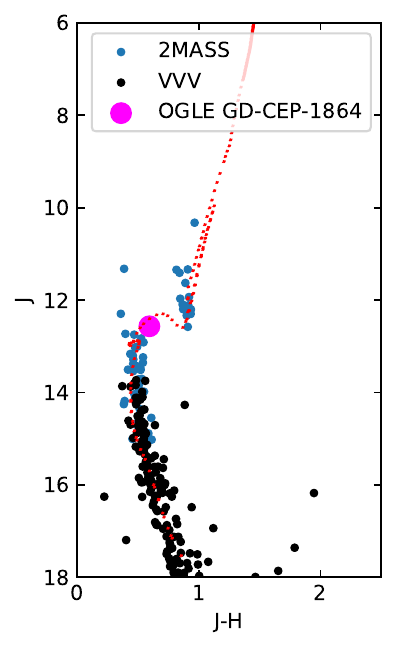} 
  \caption{VVV (black, $r \lesssim 0.5 \arcmin$) + Gaia-cleaned 2MASS (blue, $r \lesssim 1.25 \arcmin$) color-magnitude diagram for stars along the Pismis 19 sightline. A solar isochrone ($\log{\tau}=8.80\pm0.15$) was applied using parameters inferred from the cluster's red clump stars.}
 \label{fig-vvv}
\end{center}
\end{figure}

A deeper color-magnitude diagram was constructed to unveil fainter Pismis 19 members sampled by new VVV photometry, thereby improving constraints on the evolutionary status of the cluster and OGLE GD-CEP-1864. The VVV stars were standardized to 2MASS using common stars in the field.  A $J$ vs.~$J-H$ diagram (Fig.~\ref{fig-vvv}) was constructed using VVV ($r \lesssim 0.5 \arcmin$, $n=146$) and Gaia-cleaned 2MASS ($r \lesssim 1.25 \arcmin$, $n=55$) observations, whereby the former was a multi-epoch campaign, and the latter is generally a single-epoch survey.  A Padova $\log{\tau}=8.80\pm0.15$ scaled-solar isochrone was applied \citep[see also][]{sa00,bon04},\footnote{$Z=0.019$ was assumed, however, \citet{pi98} suggest that Pismis 19 could instead be marginally subsolar.} namely by partly constraining the degenerate sample space using red clump members. Specifically, the parameters utilized were $d=2.9$ kpc and $\overline{E(J-H)}=0.40\pm0.02$, and rely on the \citet{maj16} extinction laws and the \citet{lan12} red clump absolute magnitudes.  OGLE GD-CEP-1864 lies near the tip of the turnoff (Fig.~\ref{fig-vvv}).  The author C.~B.~conducted a separate $JK_s$ color-magnitude analysis following precepts outlined by \citet{bo19}, and established a comparable age of $\tau=750\pm25$ Myr ($\log{\tau}=8.87$), and nearer distance ($2.73^{+0.24}_{-0.02}$ kpc).  The determinations overlap with the red clump results within the uncertainties.

\subsection{OGLE GD-CEP-1864}
\label{s-dsct}
\begin{figure}[t]
\begin{center}
 \includegraphics[width=3.4in]{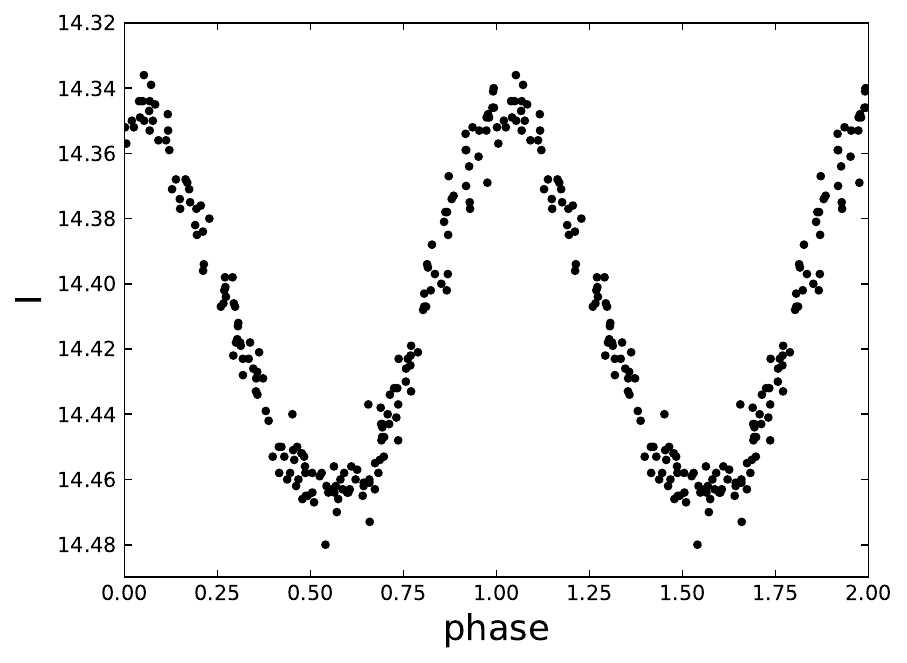} 
  \caption{$I$-band lightcurve for OGLE GD-CEP-1864 from \citet{so20}. The data were repeated beyond phase $>1$.}
 \label{fig-oglestar}
\end{center}
\end{figure}

OGLE GD-CEP-1864 was issued a preliminary classification on the basis of its light curve, which is conducive to a first overtone classical Cepheid \citep[][see also Fig.~\ref{fig-oglestar} and Table~\ref{table:ogle}]{so20}. \citet{so08} argued that first overtone classical Cepheids and $\delta$ Scutis obey the same Wesenheit Leavitt Law \citep[][]{tsv85,fer92}, and adopted a convention whereby first overtone classical Cepheids exhibit $P \gtrsim 0^{\rm d}.24$, and shortward of that period lie first overtone $\delta$ Scutis. Conversely, \citet{maj11dsct} advocated that a break in slope is discernible near $\simeq 0^{\rm d}.5$ which may differentiate the classes (their Fig.~3, which was emphasized in part thanks to new \citealt{pol10} $\delta$ Scuti photometry).\footnote{Research by Bono \& collaborators indicate the variables obey separate mass-luminosity relations \citep[e.g.,][]{bo97}.}  \citet{so23} revised their earlier interpretation after examining new observations, and reaffirmed the aforementioned break near $0^{\rm d}.5$ (bottom left panel of their Fig.~5), while still favoring that first overtone $\delta$ Scutis possess $P \lesssim 0^{\rm d}.23$ \citep[see also][]{so22}.  A separate interpretation is advocated here, namely that the division between the variable classes ($\simeq 0^{\rm d}.5$ rather than $\simeq 0^{\rm d}.23$) is based on the period linked to the slope change, \textit{and} a notable shift in the principal crossing of the instability strip. First overtone $\delta$ Scutis are just prior to the period break, whereas beyond that threshold first overtone classical Cepheids are primarily observed during later crossings of the instability strip. Importantly, \citet[][their Figs.~9 and 10]{ri22} demonstrated a distinct shift longward of the slope divergence ($0.58\pm0^{\rm d}.1$) to blue loop crossings, however, they assumed variables near opposing sides of the break are all classical Cepheids \citep[see also][]{so23}.  Fig.~13 of \citet{bo24} contains a sample of OGLE designated LMC classical Cepheids, and the later blue loop crossing does not extend to the temperatures of the faintest variables in their diagram (ostensibly $\delta$ Scutis), especially at solar metallicities \citep[for broader discussions see also][]{xu04,bo05,tu06,and16}.

\begin{deluxetable}{ccccc}
\tablecaption{OGLE GD-CEP-1864 Parameters.\label{table:ogle}}
\tablehead{\colhead{P (d)} & \colhead{I} & \colhead{I$_{a}$} & \colhead{R$_{21}$} & \colhead{$\phi_{21}$}}
\startdata
0.2919180 & 14.407 & 0.118 & 0.089 & 3.982 \\
\enddata
\tablenotetext{ }{Notes:~variable star parameters stem from \citet{so20}, and the columns represent the pulsation period, mean I-band, I-band amplitude, and Fourier coefficients.}
\end{deluxetable}

\begin{figure}[t]
\begin{center}
 \includegraphics[width=3.4in]{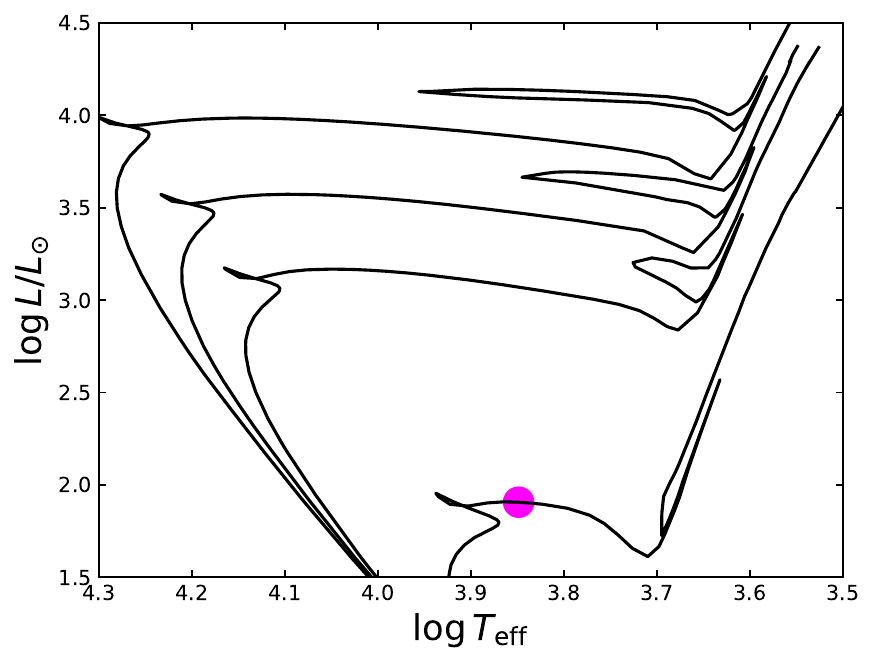} 
  \caption{An HR-diagram hosting various isochrones ($\log{\tau}=7.50, 7.75, 8.00, 8.80$) reaffirms that the blue loop does not extend to a sufficient temperature to sample OGLE GD-CEP-1864 (magenta). Scaled-solar Padova isochrones were utilized \citep[see also][]{sa00,bon04}.}
 \label{fig-hr}
\end{center}
\end{figure}

In sum, the boundary between the classes could be set by the slope discontinuity and strip crossing.  Figs.~\ref{fig-vvv} and \ref{fig-hr} indicate OGLE GD-CEP-1864 is not sampled by a canonical blue loop, which in tandem with its position prior to the period associated with the diverging slope ($\simeq 0^{\rm d}.5$): leads to a suggested reclassification as a $\delta$ Scuti variable.\footnote{See also evolved $\delta$ Scutis in NGC 1817 \citep[][the latter's Fig.~5]{ar05,maj11dsct2} and NGC 1846 \citep[][their Fig.~6]{sa18}.} Shell hydrogen burning first crossing classical Cepheids are rare \citep[e.g., Polaris,][]{tu09}, and \citet[][their Fig.~4]{tu06} argue the 2$^{\rm{nd}}$ and 3$^{\rm{rd}}$ crossings are dominant (longer) for Cepheids featuring $\log{P} \gtrsim +0.5$, whereas at lower periods the 4$^{\rm{th}}$ and 5$^{\rm{th}}$ crossings are probable.

Lastly, for a first overtone $\delta$ Scuti a Wesenheit Leavitt law tied to OGLE $VI$ LMC and Galactic photometry results in a distance of $\simeq 2.8$ kpc \citep[relation from][and $\mu_{0,LMC}$ from \citealt{st20}]{pol10}. That matches the red clump finding, however, precise knowledge of shorter-wavelength extinction laws is especially important along this sightline owing to sizable obscuration, hence the reliance on IR observations.

\section{Conclusions}
A new near-IR red clump distance for Pismis 19 reveals that DR3 parallaxes for cluster members are too small (Table~\ref{table:distances}), and possibly require a correction beyond the prescription detailed by \citet{lin21}.  OGLE GD-CEP-1864 is confirmed as a cluster member on the basis of astrometry (Fig.~\ref{fig-pm}), and its color-magnitude diagram position (Fig.~\ref{fig-vvv}).  OGLE GD-CEP-1864 ($0^{\rm d}.3$) is suggested to be a $\delta$ Scuti variable (likely first overtone), rather than a classical Cepheid (\S \ref{s-dsct}). Importantly, the case reveals that those two variable classes may be distinguished by a change in the Wesenheit slope ($\simeq 0^{\rm d}.5$) and their primary instability strip crossing.  

Going forward, continued theoretical and asteroseismology research are desirable to better characterize $\delta$ Scuti pulsations in pre-main-sequence, main-sequence, and post-main-sequence stars \citep[][]{zw11,zw17,gu21}. Moreover, a dedicated characterization of other variable stars in the field may likewise provide pertinent context, and a period-change analysis for OGLE GD-CEP-1864 could confirm a positive period increase as indicated by Fig.~\ref{fig-hr}, assuming the variable is indeed beyond the earlier hotward Kelvin-Helmholtz phase. Regarding the cluster, a future aspect to examine is whether Pismis 19 exhibits evidence of tidal stripping by the Galactic field, especially given the cluster's age ($\log{\tau}=8.80\pm0.15$, Fig.~\ref{fig-vvv}).  Examples include the broader field of NGC 6216 \citep{mt24b}, and the open cluster triad of M25, NGC6716, Cr 394 \citep{maj24triad}. 

\begin{acknowledgments}
\textbf{Acknowledgments}:~the referee's suggestions to improve the manuscript were appreciated.  This research relied on initiatives such as OGLE (\citeauthor{so08,pol10}), VMC (\citeauthor{ri22}), IRSF, CDS, NASA ADS, arXiv, Gaia, Hipparcos, and VVV (data from the ESO Public Survey program IDs 179.B-2002 and 198.B-2004, taken with the VISTA telescope, and data products from the Cambridge Astronomical Survey Unit). Dante M. gratefully acknowledges support from the Center for Astrophysics and Associated Technologies (CATA) by ANID BASAL projects ACE210002 and FB210003, and Fondecyt Project No.~1220724.  R.~K.~S.~acknowledges support from CNPq/Brazil through projects 308298/2022-5 and 421034/2023-8.  G.~Bono and V.~Braga acknowledge support from project PRIN MUR 2022 (code 2022ARWP9C), “Early Formation and Evolution of Bulge and HalO (EFEBHO)” (PI:~M.~Marconi), funded by the European Union—Next Generation EU; and large grant INAF 2023 MOVIE (PI:~M.~Marconi).
\end{acknowledgments}

\bibliography{article}{}
\bibliographystyle{aasjournal}
\end{document}